\pdfoutput=1 
\documentclass[twocolumn]{aastex631}
\usepackage{amssymb}
\usepackage{caption}
\usepackage{subcaption}

\received{October 17, 2023}
\accepted{February 20, 2024}
\submitjournal{ApJ}

\begin{document}

\title{Multiple Patchy Cloud Layers in the Planetary Mass Object SIMP0136+0933}
\correspondingauthor{Allison M. McCarthy}
\email{alliemc@bu.edu}

\author[0000-0003-2015-5029]{Allison M. McCarthy}
\affiliation{Department of Astronomy \& The Institute for Astrophysical Research, Boston University, 725 Commonwealth Ave., Boston, MA 02215, USA}

\author[0000-0002-0638-8822]{Philip S. Muirhead}
\affiliation{Department of Astronomy \& The Institute for Astrophysical Research, Boston University, 725 Commonwealth Ave., Boston, MA 02215, USA}

\author[0000-0003-2171-5083]{Patrick Tamburo}
\affiliation{Center for Astrophysics $\vert$ Harvard \& Smithsonian, 60 Garden Street, Cambridge, MA 02138, USA}

\author[0000-0003-0489-1528]{Johanna M. Vos}
\affiliation{School of Physics, Trinity College Dublin, The University of Dublin, Dublin 2, Ireland}
\affiliation{Department of Astrophysics, American Museum of Natural History, New York, NY 10024, USA}

\author[0000-0002-4404-0456]{Caroline V. Morley}
\affiliation{Department of Astronomy, University of Texas at Austin, Austin, TX 78712, USA}

\author[0000-0001-6251-0573]{Jacqueline Faherty}
\affiliation{Department of Astrophysics, American Museum of Natural History, New York, NY 10024, USA}

\author[0000-0001-8170-7072]{Daniella C. Bardalez Gagliuffi}
\affiliation{Department of Physics \& Astronomy, Amherst College, 25 East Drive, Amherst, MA 01003, USA}
\affiliation{Department of Astrophysics, American Museum of Natural History, New York, NY 10024, USA}

\author[0000-0002-0802-9145]{Eric Agol}
\affiliation{Department of Astronomy \& Virtual Planetary Laboratory, University of Washington, Box 351580, U.W., Seattle, WA 98195, USA}

\author[0000-0002-9807-5435]{Christopher Theissen}
\affiliation{Department of Astronomy \& Astrophysics, University of California, San Diego, 9500 Gilman Drive, La Jolla, CA 92093-0424, USA}

\begin{abstract}

Multi-wavelength photometry of brown dwarfs and planetary-mass objects provides insight into their atmospheres and cloud layers. We present near-simultaneous \textit{J-} and $K_s$\textit{-}band multi-wavelength observations of the highly variable T2.5 planetary-mass object, SIMP J013656.5+093347. We reanalyze observations acquired over a single night in 2015 using a recently developed data reduction pipeline. For the first time, we detect a phase shift between \textit{J-} and $K_s$\textit{-}band light curves, which we measure to be $39.9^{\circ +3.6}_{ -1.1}$. Previously, phase shifts between near-infrared and mid-infrared observations of this object were detected and attributed to probing different depths of the atmosphere, and thus different cloud layers. Using the Sonora Bobcat models, we expand on this idea to show that at least two different patchy cloud layers must be present to explain the measured phase shift. Our results are generally consistent with recent atmospheric retrievals of this object and other similar L/T transition objects.

\end{abstract}


\section{Introduction} \label{sec:intro}

Photometric variability monitoring is a useful tool for understanding the atmospheric structure of brown dwarfs, planetary-mass objects, and exoplanets. The observed variability is attributed to inhomogeneous cloud cover \citep{2014ApJ...797..120R}, thermochemical instabilities \citep{Tremblin2016}, temperature fluctuations\citep{2014ApJ...785..158R},  and/or auroral activity \citep{Hallinan2015}. Cloud structures typically persist for longer than one rotation \citep{2015ApJ...799..154M, 2017Sci...357..683A}, and as a result, we can probe the presence of these clouds by measuring the brightness of a planetary-mass object as it rotates. Past variability surveys in the infrared revealed trends with spectral type \citep{2014ApJ...797..120R,2015ApJ...799..154M}, gravity \citep{2019arXiv190306691V,2022ApJ...924...68V}, and inclination \citep{2017ApJ...842...78V}. The objects within the L/T transition, spanning from late-L to early-T, are known for having the largest variability amplitude \citep[e.g.][]{2014ApJ...797..120R,2015ApJ...799..154M}. Recently, \cite{Ashraf2022} found that variable objects can be detected by peculiarities in time-integrated spectra. 

While single-wavelength variability monitoring observations have revealed a large population of variable brown dwarfs and planetary-mass objects, multi-wavelength monitoring allows us to study time-dependent effects across the vertical structure of layers within the atmosphere. Additionally, simultaneous multiwavelength photometry probes different depths (and thus pressure/temperature levels) in the atmosphere providing insight into the different cloud layers. The vertical cloud structure can reveal itself in phase-shift and color changes across different band-passes. For example, \cite{Buenzli2012} found that for the T6.5 dwarf 2MASS J22282889-431026, phase lag increased for wavelengths probing decreasing pressure levels. Using ground-based simultaneous multi-wavelength photometric observations \cite{Biller2013} found that there were pressure-dependent phase shifts for the T0.5 object Luhman 16B \citep{2013ApJ...767L...1L,2013A&A...555L...5G}. 
 \cite{Biller2018} detected phase offsets ranging from 200$^{\circ}$ to 210$^{\circ}$ between synthesized near-IR light curves and a Spitzer mid-IR light curve for the L7 planetary mass object PSO J318.5-22. \cite{Manjavacas2019} found a tentative phase shift between \textit{J-} and \textit{H-}band light curves for the planetary mass T8 dwarf Ross 458 C. 

In this paper we investigate the T2.5 dwarf SIMP J013656.5+093347 (SIMP0136 hereafter).  SIMP0136 was discovered by \cite{2006ApJ...651L..57A} and is believed to be a member of the 200-Myr-old Carina near-moving group \citep{2017ApJ...841L...1G}. \cite{2017ApJ...841L...1G} estimated that SIMP0136 has a mass of $12.7 \pm 1.0  M_{\rm Jup}$, an effective temperature of $1098 \pm 6$ K, and a $\log g= 4.31 \pm 0.03$. In \citeyear{2009ApJ...701.1534A}, \citeauthor{2009ApJ...701.1534A} found a rotation period of $\sim$2.4 hours, a peak-to-peak \textit{J-}band amplitude of $\sim$ 50 mmag, and reported light curve evolution from night-to-night. \cite{Apai2013}, \cite{2013AN....334...40M}, and \cite{Croll} each found a rotation period consistent with \cite{2009ApJ...701.1534A}'s original $\sim$ 2.4-hour rotation period.
\cite{2016ApJ...818...24K} reported highly circularly polarized pulsed radio emission in the 4-8 GHz band and a magnetic field $\gtrsim 2.5$ kG, consistent with the presence of aurorae on SIMP0136. The authors adopted a period of $2.88 ^{+0.34}_{-0.27}$ hr for the quasi-quiescent emission at \textit{X-} band. A detailed investigation into the inclination angle impacting observed color and variability found SIMP0136 to be close to equator-on at $80\pm 12 ^{\circ}$\citep{2017ApJ...842...78V}.

\citet{2009ApJ...701.1534A} also completed consecutive multi-wavelength observations of SIMP0136 which revealed correlated \textit{J-} and $K_s$\textit{-}band light curves with a $\frac{\Delta Ks}{\Delta J}$~=~0.48~$\pm$~0.06. The authors showed that a two-temperature surface could not explain the$\frac{\Delta Ks}{\Delta J}$ value, and as a result they interpreted the variability results as clouds. \cite{Apai2013} used the Hubble Space Telescope (HST) WFC3 grism mode to measure spectroscopic variability and found that there were minimal \textit{J-H} color changes. The authors also determined that the surface is covered by a spatially heterogeneous mix of two distinct regions: low-brightness, low-temperature thick clouds, and brighter, thin, and warm clouds. \cite{2020AJ....159..125L} also reported a lack of \textit{J-H} color change. No specific values are cited in the text of the paper, but based on the linear fit shown in Figure 5, the $\Delta (J-H)$ appears to be 0, with minimal error. \cite{2016ApJ...826....8Y} reported a phase shift of $33.4 \pm 3.9 ^{\circ}$ between simultaneous Spitzer 3.6 $\mu m$ and HST light curves. Additionally, the authors report a range of phase shifts between Spitzer 3.6 $\mu m$ and 4.5 $\mu m$ bands. 

Using the Brewster retrieval framework and archival 1\textendash 15 $\mu m$ spectra, \cite{Vos2022} completed an atmospheric retrieval of SIMP0136. Building on the work of \cite{Apai2013}, they found that a patchy, high-altitude forsterite (Mg$_2$SiO$_4$) cloud above a deeper, optically thick iron cloud best describes their spectral data. The authors determine that in order to reproduce the  \cite{Apai2013} HST variability amplitude observed in SIMP0136, the forsterite cloud coverage changes by a small amount as the object rotates. Based on an amplitude variability of 3\%, \cite{Vos2022} find that the forsterite cloud coverage fluctuates between 69\% and 72\%.

In this paper, we report the first observed phase shift between \textit{J-} and $K_s$\textit{-}bands for SIMP0136. We reanalyze near-simultaneous \textit{J-} and $K_s$\textit{-}band observations, originally taken in November 2015 using the 1.8-m Perkins Telescope Observatory near Flagstaff, Arizona, and presented in \cite{Croll}, with the newly developed \texttt{PINES Analysis Toolkit} \citep[\texttt{PAT,}][]{PINES_analysis_toolkit}. We use the Sonora Bobcat (cloud-free) models \citep{Sonora} to calculate the pressure/temperature levels probed by \textit{J-} and $K_s$\textit{-}band. We find that \textit{J-} and $K_s$\textit{-}bands probe different regions of the atmosphere and thus would be affected differently by the presence of different cloud levels. We find that to explain the observed phase shifts, we must invoke the presence of at least two different patchy cloud layers. In Section 2, we discuss our observations and data reduction technique. In Section 3, we describe our analysis and present our results, followed by our prediction of cloud distribution in Section 4. In Section 5, we use a three-patch atmospheric model to explain the observed color changes of SIMP0136. In Section 6, we present our conclusions and future steps.

\section{Observations and Data Analysis} \label{sec:Observations}

The data presented in this paper were originally taken on UT 20 November 2015 as part of a larger campaign reported in \cite{Croll}, which describes the observations and sky conditions. Briefly, the observations were taken using the Mimir near-infrared instrument \citep{2007PASP..119.1385C} at the 1.8-m Perkins Telescope Observatory, located on Anderson Mesa near Flagstaff, Arizona. The data consist of \textit{J-} and $K_s$\textit{-}band exposures taken sequentially with 12, 30-second \textit{J-}band exposures, followed by 24, 15-second $K_s$\textit{-}band exposures. The overhead time, including time for the readout, was $\sim$ 3 seconds for each observation, regardless of the band. This pattern continued with minimal interruption for 8.5 hours. The guiding was performed by an off-axis guide camera. The observations were taken in photometric conditions. Calibration images were also obtained during the November 2015 run. Flat fields were obtained using a lamp-illuminated dome flat screen. One hundred fifty lamp-on and lamp-off images were taken in both \textit{J-} and \textit{$K_s$}-bands. Twenty 30-second darks and 20 15-second darks were also taken. 

We independently reduced the raw data using the newly developed \texttt{PINES Analysis Toolkit} \citep[\texttt{PAT,}][]{PINES_analysis_toolkit}. \texttt{PAT} was originally designed to reduce data from the Perkins INfrared Exosatellite Survey, PINES \citep{PINES1}. \texttt{PAT} has proven to be applicable, more broadly, to any photometric observations taken using the Mimir Instrument at the Perkins Telescope Observatory. \texttt{PAT} is explained in detail in Section 3 of \cite{PINES1}, but the procedure will be explained briefly here. The following procedure is completed on the \textit{J-} and \textit{$K_s$-}band data independently.

\texttt{PAT} begins by combining the calibration images and measuring a bad pixel mask, including the hot, dark, and variable pixels on the Mimir detector. The data are reduced using standard calibration techniques. The user then selects an image which the pipeline then analyzes to find the target and suitable reference stars using source detection procedures from the \texttt{photutils} Python package. The user confirms this information before proceeding. This step is repeated for each band. \texttt{PAT} then measures the centroid location of the target and reference stars in each exposure. Next, the pipeline performs simple aperture photometry with both fixed and time-variable apertures. Throughout this process, if a bad pixel (identified by the bad pixel mask) falls on either the target or reference star, the pipeline uses a 2D-Gaussian to interpolate over the bad pixel. Finally, \texttt{PAT} produces light curves by using the background-subtracted flux from the target and reference stars. \texttt{PAT} also creates an ``artificial light curve", or ALC, a unitless weighted sum of normalized reference star fluxes, following the approach of \cite{2020MNRAS.495.2446M}, to remove trends due to changes in the transmission in the Earth's atmosphere and weight the individual reference stars by their signal-to-noise ratio and intrinsic variability. The pipeline then selects the optimal aperture (fixed or time-variable) by selecting the light curve that minimizes the average scatter over the duration of individual blocks of data. In this context, the term \textit{block} refers to one $\sim$6 minute set of observations before the filter is switched. In the case of our data, the optimal aperture for \textit{J-}band was found to be fixed at a {2}\farcs{9} radius (5 pixels), and a fixed {1}\farcs{7} radius (3 pixels) for $K_s$\textit{-}band. The high background present in $K_s$\textit{-}band results in increased photon noise for larger apertures. 

In order to obtain near-simultaneous \textit{J-} and $K_s$\textit{-}band observations, the filter was changed every $\sim$6 minutes. We measured the mean flux and error on the mean flux for each of these 6-minute blocks of data, resulting in 37 binned points in \textit{J-}band and 36 binned points in \textit{$K_s$-}band. Figure ~\ref{fig:JandKindlc} shows the resulting light curves.  The black markers represent the binned data, and the small, light markers --- orange dots for \textit{J-}band, and pink x's for $K_s$\textit{-}band --- represent the unbinned data.

Since \texttt{PAT} reduces the \textit{J-} and \textit{$K_s$-}band data independently,the reference stars used in the reduction are not necessarily the same in both \textit{J-} and \textit{$K_s$-}band. As such, the reference stars used in the reduction only need to have similar \textit{J-} \textit{or} \textit{$K_s$-}band magnitudes as SIMP0136, not necessarily the same \textit{$J-K_s$} color. The ALC is best used to measure changes in magnitudes, not the magnitudes themselves. 

We also explored the impact of precipitable water vapor on the measured flux variations of SIMP0136 and the reference stars in each band. Figure 8 in \cite{PINES1} shows the impact of precipitable water vapor for \textit{J-band}. It shows that for a change in 1mm of precipitable water vapor, there is a 0.1\%, or 1mmag, of variation, which is well within our measured variability of SIMP0136. The \textit{$K_s$-band} analysis was done in the same way, and found a 0.4\%, or 4mmag, of variation, which is also within our measured variability of SIMP0136. Additionally, we determined that there was no trend between temperature of the reference star and increased variability in the object. We concluded that the impact of precipitable water vapor is negligible.

\section{Comparing the \textit{J-} and \textit{$K_s$}-band Light Curves}

\subsection{MCMC Modeling}

\begin{figure*}
\centering
\includegraphics[width=0.99\textwidth]{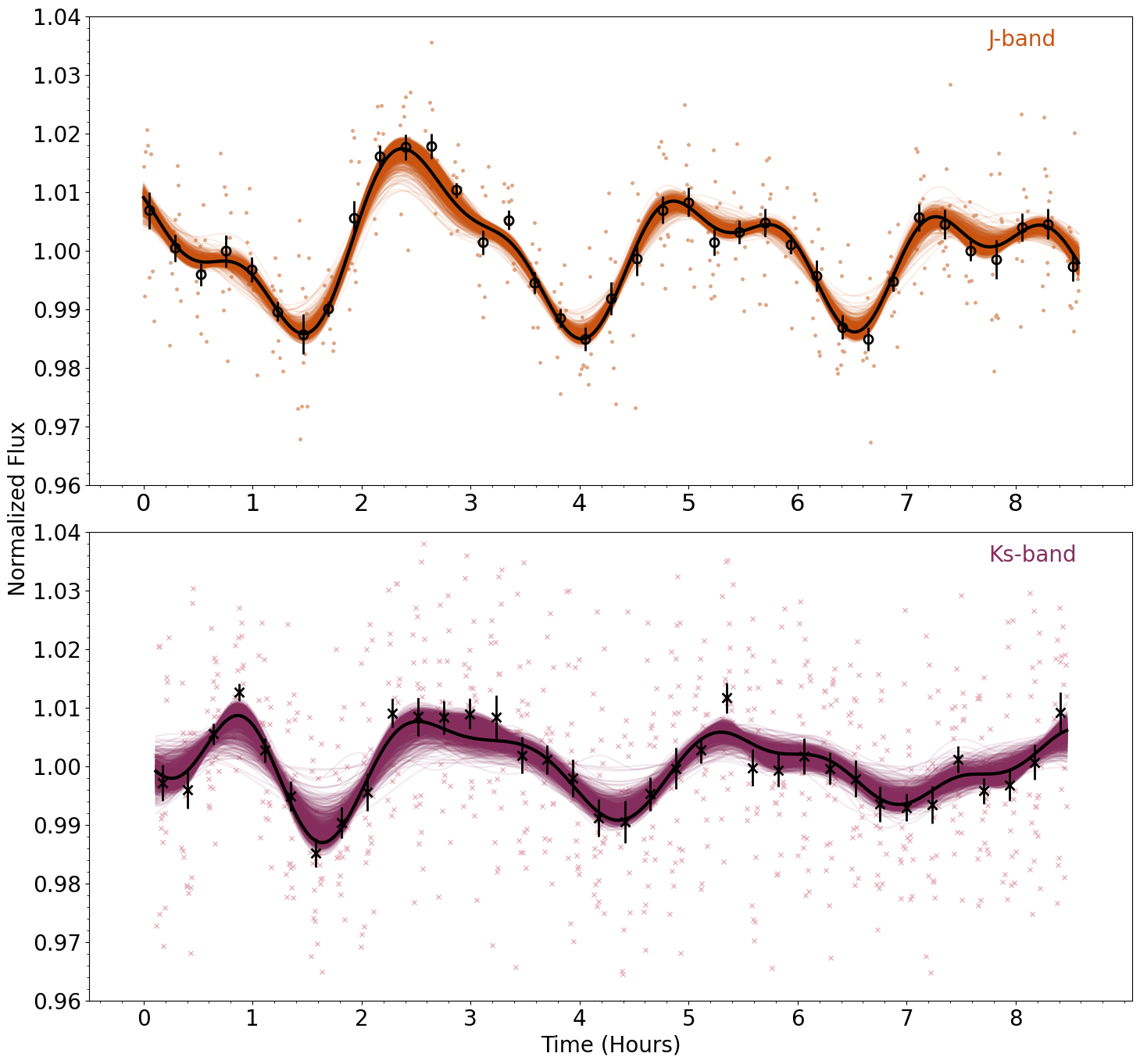}
\caption{\textit{J-} and $K_s$\textit{-}band light curves of SIMP0136. The \textit{J-}band light curve is shown above in orange with round markers. The $K_s$\textit{-}band light curve is shown below in purple with x markers. The small, light markers represent the unbinned data. The large, black markers are the binned data with 1 $\sigma$ error bars, calculated using the standard deviation of the mean within a block. The orange and purple lines show 1000 randomly selected MCMC samples for the model light curves of \textit{J-} and $K_s$\textit{-}band, respectively. In both plots, the black line is the maximum likelihood solution of the MCMC fits.\label{fig:JandKindlc}}
\end{figure*}

In order to determine the phase shift between the \textit{J-} and $K_s$\textit{-}band light curves, we chose to cross-correlate the light curves. The observing strategy results in the data being segmented into blocks and gaps. This segmentation poses a challenge when calculating the phase shift between the \textit{J-} and $K_s$\textit{-}band light curves, as it is not possible to cross-correlate data with different x-axis timestamps. To address this issue, we chose to obtain a best-fit model for the data and to use the model to interpolate the measurements onto the same time stamps. 

Using the reduced data from \texttt{PAT}, we used \texttt{celerite 2} \citep{celerite1,celerite2}, an algorithm for fast and scalable Gaussian Processes (GP), to determine a best-fit model. The \texttt{RotationTerm} kernel, designed for stellar rotation, uses a combination of two stochastically driven damped simple harmonic oscillator (SHO) terms. The period of the second mode is half that of the first mode's period. The model had five free parameters \textemdash sigma, period, Q0, dQ, and f \textemdash that are described fully in the \texttt{celerite 2} documentation.

To determine best-fitting parameters and uncertainties in the model, we performed a Markov chain Monte Carlo (MCMC) using the \texttt{emcee} package \citep{2013PASP..125..306F}. We chose a uniform prior for the period. The limits set were between $\sim$0.9 and $\sim$3.6 hours based on previous measurements of the system. 

We ran 64 chains with 6,000 steps, with the first 1,000 discarded as burn-in. The chains were well mixed by visual inspection. Figure \ref{fig:JandKindlc} shows the results of the MCMC as orange lines for \textit{J-}band and purple lines for $K_s$\textit{-}band, while the black line shows the maximum likelihood solution of the MCMCs in both plots. We also inspected the median posterior solution, which was nearly identical.

\subsection{\textit{J/}$K_s$ Phase Shift}

\begin{figure*}
\centering
\includegraphics[width=0.95\textwidth]{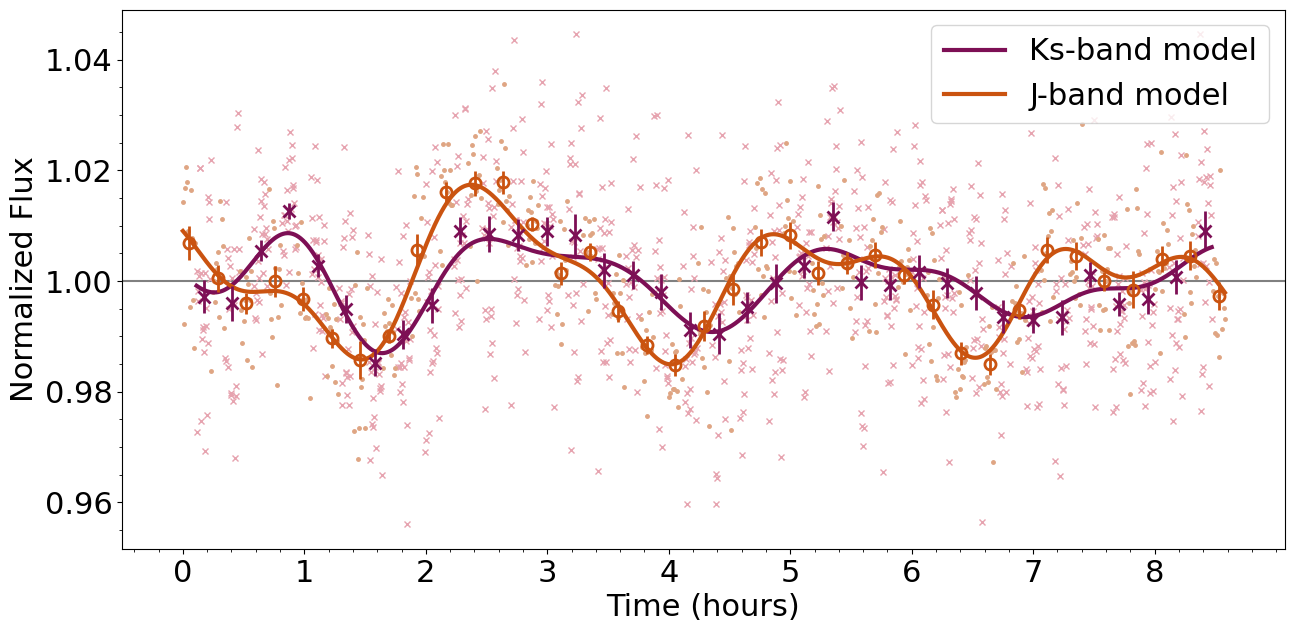}
\caption{Maximum likelihood solution of the MCMC fits for the \textit{J-} and $K_s$\textit{-}band data. The \textit{J-}band data is shown in orange, with round markers, and the $K_s$\textit{-}band data is shown in purple, with x markers. The small, light markers are the unbinned data.\label{fig:JandKLC}}
\end{figure*}

Figure \ref{fig:JandKLC} shows the \textit{J} and \textit{$K_s$} data and best-fit models on the same plot, revealing similar features that are slightly out of phase. To quantify the phase shift between the two light curves, we used the \texttt{signal} function in the \texttt{scipy} Python package \citep{SciPy-NMeth2020} to perform a cross-correlation analysis on the maximum likelihood solution for each light curve. 

\begin{figure*}
  \centering

  \begin{subfigure}{0.48\textwidth}
    \centering
    \includegraphics[width=\linewidth]{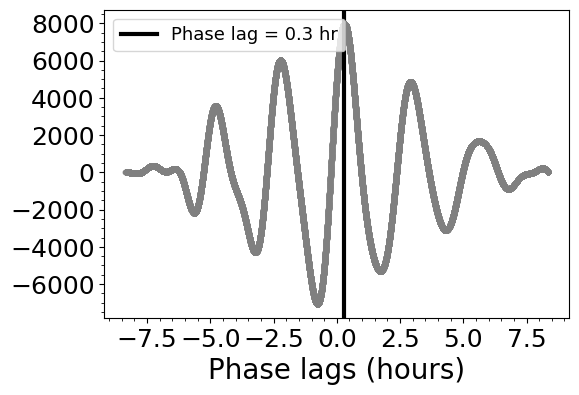}
    \caption{}
    \label{fig:CrossCorr}
  \end{subfigure}
  \hfill
  \begin{subfigure}{0.48\textwidth}
    \centering
    \includegraphics[width=\linewidth]{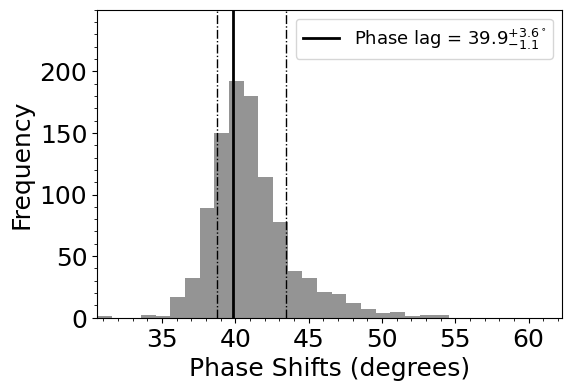}
    \caption{}
    \label{fig:ShiftDistrib}
  \end{subfigure}

  \caption{Panel (a) shows the cross correlation results of the \textit{J-} and $K_s$\textit{-}band light curve MCMC model predictions. The phase shift is found to be 0.3 hr = $39.9^{\circ +3.6}_{-1.1}$. Panel (b) shows the distribution of phase shifts using the cross-correlation technique between 1000 random model fits for each \textit{J-} and $K_s$\textit{-}band. We determined the errors on the phase shift by calculating the 16th and 84th percentile of the data, shown in dashed lines. Since the distribution of phase shifts is right-skewed, this led to a larger positive error than a negative error: $39.9^{\circ +3.6}_{-1.1}$. }
  
\end{figure*}

To estimate the error on the cross-correlation, we randomly selected a chain from each of the \textit{J-} and $K_s$\textit{-}band MCMC fits, and performed the cross-correlation on those two randomly selected chains. We repeated this process 1000 times. The distribution of the phase shifts was right-skewed as shown in Figure ~\ref{fig:ShiftDistrib}. Because of this, we report the 16th and 84th percentiles for the lower and upper errors respectively. The result of this analysis is presented in Figure \ref{fig:CrossCorr}, which shows a $39.9^{\circ +3.6}_{-1.1}$ phase shift between J and $K_s$\textit{-}band for SIMP 0136.

\subsection{\textit{J} and \textit{$K_s$} Period and Amplitude}

From the results of the MCMC, we find a period of P=$2.55^{+0.10}_{-0.08}$ hours for \textit{J-}band. For $K_s$\textit{-}band, we find a period of P= $2.68 \pm 0.28$ hours. The two period values are consistent to within 1$\sigma$. 

Following the approach of \cite{Apai2013}, we also measured the time between peaks and troughs for both light curves. We used the same method to determine the uncertainties on time between peaks as the phase shift. In both light curves, but especially prominent in \textit{J-}band, is a double peak feature. For this study, we measured the time between the first peaks in each of the double peak features.

For \textit{J-}band, between the first and second troughs, $t_{t12}=2.55 \pm 0.02$ hr, and between the second and third troughs, $t_{t23}=2.52 \pm 0.01$ hr. Between the first and second peaks, $t_{p12}=2.49 \pm 0.02$ hr, and between the second and third peaks, $t_{p23}=2.40 \pm 0.03$ hr. 

For $K_s$\textit{-}band, between the first and second troughs, $t_{t12}=2.70 \pm 0.05$ hr, and between the second and third troughs, $t_{t23}=2.60 \pm 0.05$ hr. For the peaks in $K_s$\textit{-}band, we only calculated one difference for the peaks between $\sim 5.19$ and $\sim 2.41$ hr, $t_{p12}=2.76 \pm 0.07$ hr.  

In both \textit{J-} and $K_s$\textit{-}band, the amplitude of the light curve decreases over the course of the observations. For \textit{J-}band, we begin by measuring the amplitude from the trough at $\sim 1.47$ hr to the peak at $\sim 2.37$ hr, and find the amplitude is $3.13 \pm 0.16\%$. In the next rotation, the \textit{J-}band amplitude decreases to $2.27 \pm 0.10\%$. Finally, in the third rotation it decreases again to $1.93 \pm 0.12\%$. For $K_s$\textit{-}band, beginning at the peak at $\sim 0.75$ hr and measuring to the trough at $\sim 1.53$ hr, the amplitude is $1.94 \pm 0.27\%$. In the second rotation, the amplitude decreases to $1.57 \pm 0.19\%$. In the final rotation, the amplitude decreases again to $1.17 \pm 0.16\%$. 

The change in amplitude throughout the night, as well as difference in timing between peaks in the light curve, combined with the phase-shift between the bands leads to color modulation throughout the course of the observation and can be indicative of a rapidly evolving atmosphere\citep[e.g.][]{2015ApJ...799..154M,2017Sci...357..683A}.

\section{Cloud Distribution Scenarios in the Atmosphere of SIMP0136} \label{sec:Discussion}
\subsection{Sonora Bobcat Models}

Simultaneous multiwavelength photometry probes different depths, and thus pressure/temperature levels in the atmosphere of brown dwarfs and planetary mass objects. These observations provide insight into the vertical cloud structure in the atmosphere of these objects. To calculate the flux contribution from different pressure levels, we used the cloudless Sonora Bobcat models \citep{Sonora, SonoraBobcat, SonoraKcoefs}. 

The Sonora Bobcat models are available for a range of atmospheres with varying effective temperatures, gravities, and metallicities. Using an atmospheric retrieval analysis of SIMP0136, \cite{Vos2022} found a [M/H] of $-0.32 \pm 0.05$ and a C/O of $0.79 \pm 0.02$. Additionally, they found a $T_{\rm eff}=1329 \pm 17$ K, which they conclude is too high, and the SED-calculated temperature would be a better estimate. The authors' SED temperature estimate is $T_{\rm eff}=1150 \pm 70$ K, which is in agreement with a previous result from \cite{2017ApJ...841L...1G} of $1098 \pm 6$ K. The estimate of the temperature found from the retrieval is determined in conjunction with the radius. The retrieval results also find a small radius: $0.81 \pm$ 0.03 $R_{Jup}$. The small radius, and associated large temperature, are not consistent with evolutionary models, but are a common feature of retrieval-derived radii \citep[e.g.][]{Burningham2021, Molliere2020,Barman2011}. Finally, \cite{Vos2022} find the log g = $4.5 \pm 0.4$ calculated from  retrievals to be similar to the log g = $4.25 \pm 0.08$ calculated from the SED fit. Using these values as a reference, and working within the available values from the Sonora Bobcat models, we used the atmospheric structure model with the following parameters: [M/H]=-0.5, solar C/O ratio, $T_{eff}$=1100 K, and g= 316 m/s (log g= 4.5). We used the correlated $k$-coefficients \citep[described fully in][]{1991JGR....96.9027L} with the following parameters: [M/H]=-0.5, and solar C/O ratio. The solar C/O ratio which Sonora Bobcat uses is from  \cite{Lodders2010} value for the solar C/O=0.458.

To begin, we downloaded the appropriate correlated $k$-coefficients from \cite{SonoraKcoefs}. We then obtained the pressure-temperature (PT) profiles from the Sonora Bobcat atmospheric structure files \citep{SonoraBobcat}. We interpolated the temperature values onto a new axis that was evenly spaced in $\Delta$ log P to create a more finely sampled profile. We then interpolated the k-coefficient grid onto the PT profile for the model atmosphere. 

\begin{figure*}
\centering
\includegraphics[width=\textwidth]{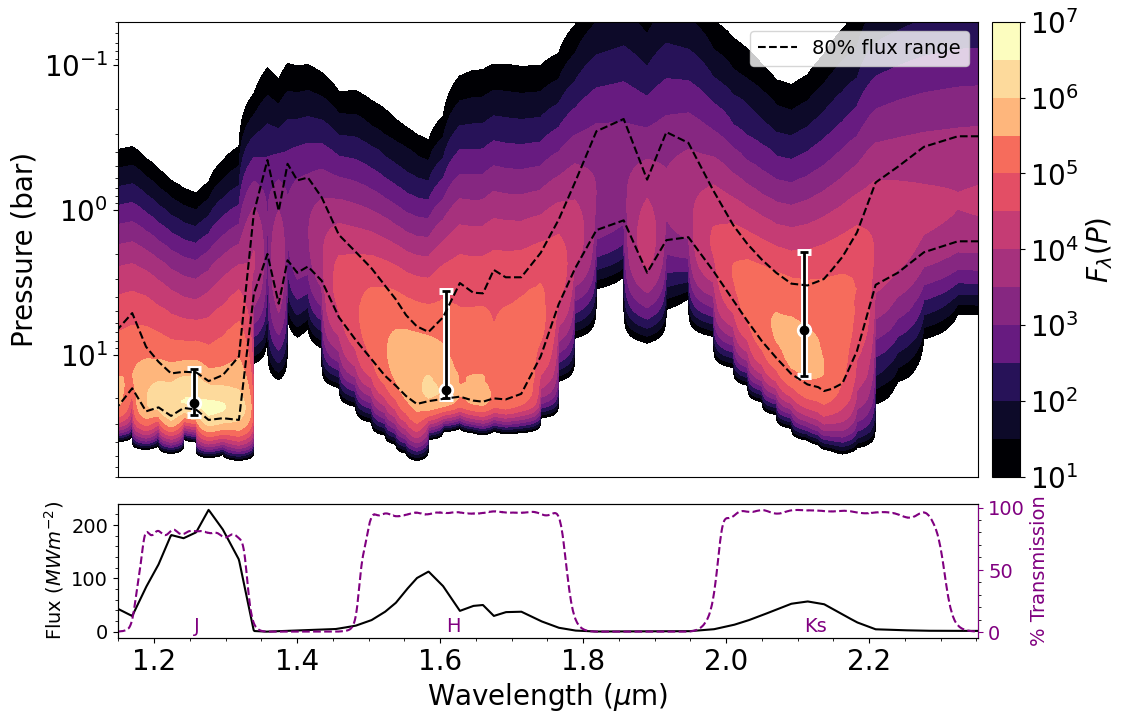}
\caption{\textbf{Top:} The contour plot shows the distribution of flux as a function of wavelength and pressure for a cloudless atmosphere with effective temperature of 1100K, log(g)=4.5, [M/H]=-0.5, and solar C/O. The color bar distribution is logarithmic. The dotted black lines represent the pressure levels of the atmosphere where 10\% and 90\% of the flux is emitted. The circle markers are located on the weighted average wavelength and peak flux emission pressure level for each bandpass while the bars show the range of pressure levels where 80\% of the emitted flux is weighted by the band transmission. \textbf{Bottom:} This plot displays the Near Infrared spectra of the Sonora Bobcat model in black, and the transmission curves for \textit{J-}, \textit{H-}, and $K_s$\textit{-}band in purple.}
\label{fig:BigPlot}
\end{figure*}

Making the assumption that the atmosphere is in local thermal equilibrium, we solved the radiative transfer equation along the PT profile to determine the flux contribution from each layer at each wavelength. The results are shown in the contour plot of Figure \ref{fig:BigPlot}. The dotted lines shown in Figure \ref{fig:BigPlot} contain the pressure levels at which 80\% of the flux originates. 

The observed data does not include \textit{H-}band measurements, but we chose to include the \textit{H-}band wavelengths in Figure \ref{fig:BigPlot} for the sake of mapping the complete near infrared wavelengths. The plotted dots and associated error bars show the pressure levels at which 80\% of the total flux within each band, \textit{J}, \textit{H}, and \textit{$K_s$}, originates \citep[similar to Figure 5 in][]{Biller2013}. To determine these values, we summed the radiative transfer equation results for each wavelength window in each bandpass to obtain the total emergent flux. Next, we multiplied the emergent flux and the transmission curve for the band. We then determined the PT levels corresponding to 10\% and 90\% of the total flux in each bandpass and calculated a cumulative summation of each PT level until the PT levels from below which the flux accounted for 10\% and 90\% of the total flux in the band were obtained. The location of the marker corresponds to the weighted average wavelength and the maximum flux contribution pressure level for each band. The plotted data points for each band do include the transmission curve in the calculations for the location of the marker and the size of the error bars. The pressures for which the center 80\% of the received (emitted flux multiplied by the transmission curve) flux originates for each band are 12.6 \textendash 26.2 bar for \textit{J-}band, 3.6 \textendash 20.0 bar for \textit{H-}band, and 1.9 \textendash 14.1 bar for $K_s$\textit{-}band.

\begin{figure}
\centering
\includegraphics[width=\linewidth]{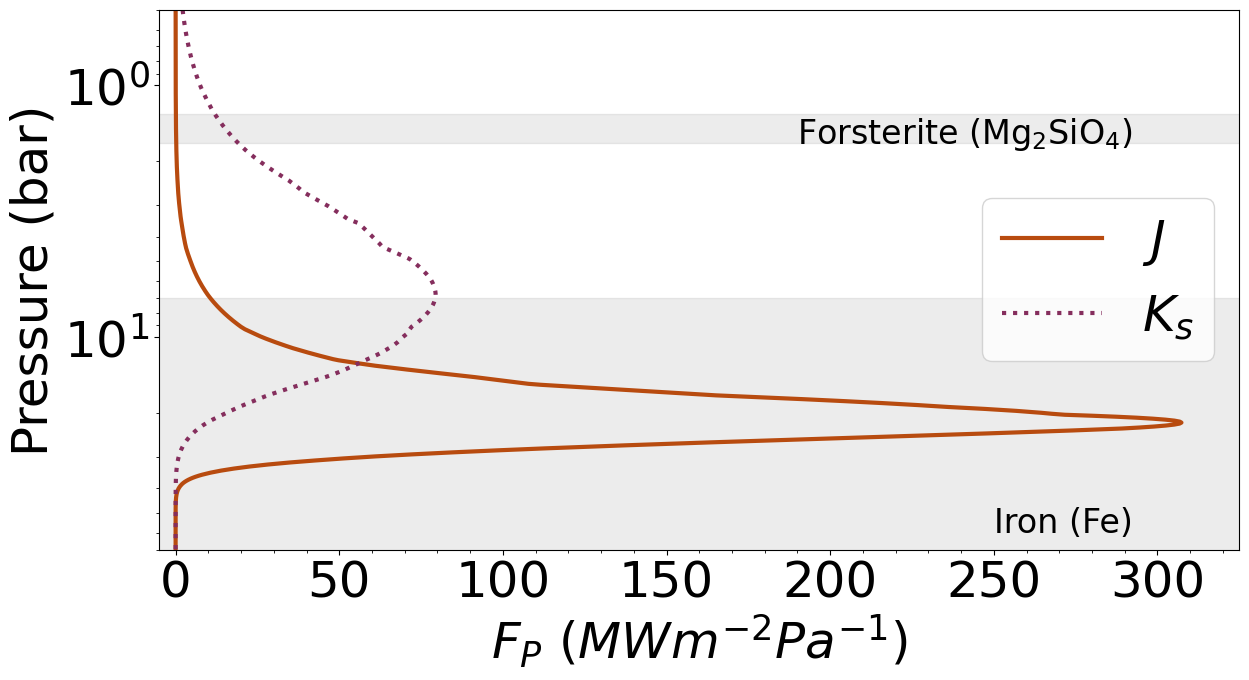}
\caption{Received flux distribution for the \textit{J-} and $K_s$\textit{-}band pass wavelength range as a function of pressure for a cloudless atmosphere. The solid orange line and dotted purple line represent the \textit{J-} and $K_s$\textit{-}band flux distribution, respectively. The gray boxes in the plot depict the regions of the atmosphere which \cite{Vos2022} found SIMP0136 contains forsterite (Mg$_2$SiO$_4$) and iron (Fe) clouds.}
\label{fig:F_P}
\end{figure}

The lower plot in Figure \ref{fig:BigPlot} displays the NIR spectrum of the Sonora Bobcat model in black and the transmission curves for the \textit{J-}, \textit{H-}, and $K_s$\textit{-}, bandpasses in purple. The flux units are in megawatts. Similarly, Figure \ref{fig:F_P} shows the total flux per pressure for \textit{J-} and $K_s$\textit{-}band with the units in megawatts. This plot highlights the different regions of the atmosphere that the \textit{J-} and $K_s$\textit{-}band probe. The gray bars represent the pressure levels at which the clouds identified by \cite{Vos2022} are situated.

\subsection{Multiple Atmospheric Layers Responsible for Phase Shift} \label{sec:Patchy Cloud Layers}

While a combination of any three unique atmospheric regions or layers could create \textit{J-} and $K_s$\textit{-}band variability, as well as the phase shift between the two light curves, previous studies such as \cite{Vos2022} have found strong evidence for the presence of clouds. Therefore, we considered different scenarios of cloud distribution in the atmosphere of SIMP0136 to determine which would result in the modulations in the \textit{J-} and $K_s$\textit{-}band, as well as the phase shift between the two light curves. 
In these scenarios, we consider clouds to be blocking a portion of the flux from the region of the atmosphere which the band probes. We acknowledge that, in reality, changing cloud opacities within or above the portion of the atmosphere that the band is probing, will alter the distribution of the emergent flux per pressure level.  

We considered four simple cloud distribution scenarios to determine what could be causing the phase shift between \textit{J-} and $K_s$\textit{-}band for SIMP0136:

\begin{enumerate}
    \item {No clouds} within or above the regions of the atmosphere which \textit{J-} and $K_s$\textit{-}band probe. In this scenario, there would be no clouds to block the emergent flux from the regions of the atmosphere that either the \textit{J-} or $K_s$\textit{-}band probe. If this were the case, we would see no modulation in either light curve.
    \item {Patchy upper atmosphere clouds} either within or above the region of the atmosphere which $K_s$\textit{-}band probes. In this scenario, the clouds in the upper atmosphere would block the emergent flux from both the \textit{J-} and $K_s$\textit{-}band regions of the atmosphere. This would create modulation in both the \textit{J-} and $K_s$\textit{-} light curves and the modulation would be in-phase.
    \item {Patchy lower atmosphere clouds} solely within the region of the atmosphere which \textit{J-}band probes. In this scenario, the emergent flux from the portion of the atmosphere that $K_s$\textit{-}band probes, would be unobstructed, and there would be no modulation in the $K_s$\textit{-}band light curve. However, the clouds would cause disruptions to the emergent flux from the region of the atmosphere which \textit{J-}band probes and the \textit{J-}band light curve would reflect this by being variable. 
    \item {Upper atmosphere either above or within the region that $K_s$\textit{-}band probes and lower atmosphere clouds within the region of the atmosphere which \textit{J-}band probes.} In this scenario, the presence of clouds in both the upper and lower atmosphere would produce variable light curves for both \textit{J-} and $K_s$\textit{-}band. However, since the lower atmosphere clouds would only affect the emergent flux from the region of the atmosphere that the \textit{J-}band probes, the light curves would likely not be in phase. The light curves could be in phase if the upper and lower atmosphere clouds were of the same size, shape, and in-phase. 
\end{enumerate}

Based on these four simple atmosphere structure scenarios for SIMP0136, we conclude that there must be at least two patchy cloud layers, one above or within the atmospheric region that $K_s$\textit{-}band probes (hereafter referred to as \textit{upper atmosphere clouds}), and one between the regions that $K_s$\textit{-} and \textit{J-}band probe or within the \textit{J-}band region (hereafter referred to as \textit{lower atmosphere clouds}), in order to produce the phase-shift seen in our data. 

These conclusions build on the work of \cite{Vos2022} \citep[which in turn builds on the work of ][]{Apai2013} who find that the spectrum of SIMP0136 is best described by a model containing patchy forsterite clouds lying above an iron cloud deck. \cite{Vos2022} show that the forsterite clouds exist at a pressure level of 1.3-1.7 bar, and the iron clouds exist at a pressure level of and below $\sim$7 bar. We used these pressures for the basis of the location of the gray bars shown in Figure \ref{fig:F_P}. The atmospheric structure suggested by \cite{Vos2022} matches scenario (2)\textemdash  a patchy upper atmosphere cloud over a ``uniform" lower cloud deck. We argue that in order to explain the phase shift, both cloud layers must be patchy, matching atmospheric structure scenario (4) discussed in Section \ref{sec:Patchy Cloud Layers}.

As shown in Figure 5 of \cite{Vos2022}, the addition of an iron cloud layer into the atmospheric model does alter the flux contribution per pressure level, especially around 1-1.3 microns. The addition of the forsterite patch raises the pressures probed in J-band to $\sim$0.6 bar. This discrepancy between the clear atmospheric model, and the Brewster retrieval model from \cite{Vos2022}, while important to consider, does not alter the fact that in order to explain the phase shift, multiple patchy cloud layers must be present. 

In order to create a phase-shift between any two curves, the addition of another curve must be present. For the atmosphere of SIMP0136, the varying cloud coverage in each cloud layer influences the observed light curve at each wavelength. When forsterite clouds are present, it follows that the optical depth of both \textit{J-} and $K_s$\textit{-}band will rise to a lower pressure level in the atmosphere. However, in the region with patches of the forsterite cloud, the J-band observations are still able to probe down to the level which the iron clouds exist, while even in a hole in the forsterite cloud layer, the $K_s$\textit{-}band observations cannot probe down to the depth of the atmosphere where iron-clouds exist. This is highlighted by our clear atmosphere model shown in Figure \ref{fig:BigPlot}. So, despite which pressure layer we are able to probe when forsterite clouds are present, the important factor is that when forsterite clouds are \textit{not} present, the J-band can probe down to the iron cloud layer, and observe its modulations, while the $K_s$\textit{-}band cannot.

\section{Implementation of Three Patch Atmospheric Model}\label{sec:modelingatmosphere}

In an attempt to model the observed phase shift, we used a combination of the Sonora Bobcat (clear atmosphere) models \citep{Sonora}, the Sonora Diamondback (cloudy atmosphere) models \citep{Morley2024Preprint}, and bespoke Sonora Diamondback iron-only cloudy models to create a three-patch model atmosphere that could describe the observed period, amplitude, and phase shift when each patch was modulated. The standard Sonora Diamondback models included five values for the sedimentation efficiency parameter \textemdash $f_{sed}$=1, $f_{sed}$=2, $f_{sed}$=3, $f_{sed}$=4, and $f_{sed}$=8, (lower $f_{sed}$=cloudier) \textemdash and the bespoke Sonora Diamondback iron-only models included four $f_{sed}$ values \textemdash $f_{sed}$=2, $f_{sed}$=3, $f_{sed}$=4, and $f_{sed}$=8. In each of the following subsections, we describe the outcomes of each different patchy atmosphere which we created. Subsection \ref{1100k} explores a three-patch model with a cloudy patch (including forsterite clouds), an iron only patch, and a clear patch all at an effective temperature of 1100K. Subsection \ref{521} considers an entirely clear atmosphere with three patches at three different effective temperatures, while subsection \ref{522} considers an entirely cloudy atmosphere with each of the three patches having three different effective temperatures. Subsections \ref{531}, \ref{532}, and \ref{533}, combine clear and cloudy atmospheric patches at different effective temperatures. In each case, our procedure was the same.

We combined three spectra, from the different models, representing the different patches of the atmospheres, at different ratios. Representing the three different patches spectra as $s_1$, $s_2$, and $s_3$, their three ratios as $a$, $b$, and $c$ (such that $a+b+c=1$), and the total combined spectra as $S$, we combine the three spectra as 
$$S=(a*s_1)+(b*s_2)+(c*s_3).$$

\begin{figure*}

\includegraphics[width=0.95\textwidth]{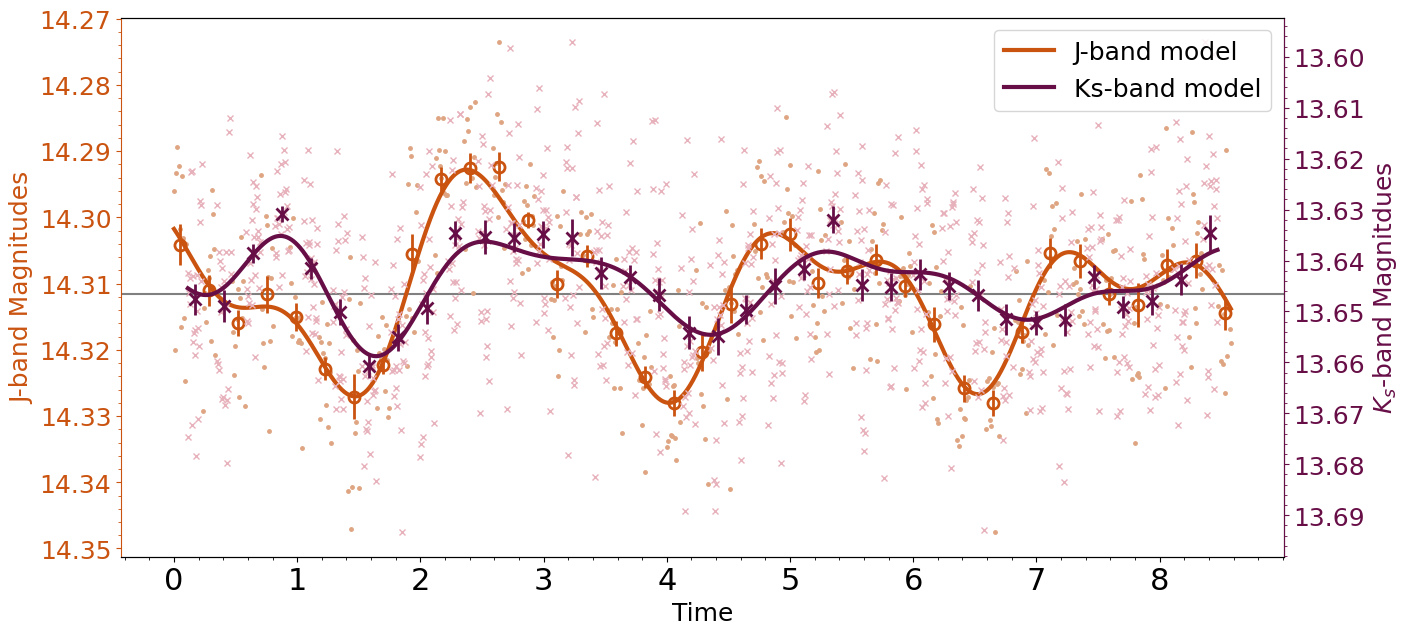}
\caption{Maximum likelihood solution of the MCMC fits for the \textit{J-} and $K_s$\textit{-}band data. The \textit{J-}band data is shown in orange, with round markers, and the $K_s$\textit{-}band data is shown in purple, with x markers. The small, light markers are the unbinned data. The \textit{J-}band magnitudes are shown on the left axis, and the $K_s$\textit{-}band magnitudes are shown on the right axis.} 
\label{fig:JandKLCMags}
\end{figure*}

\begin{figure*}
    \centering
    \includegraphics[width=0.9\textwidth]{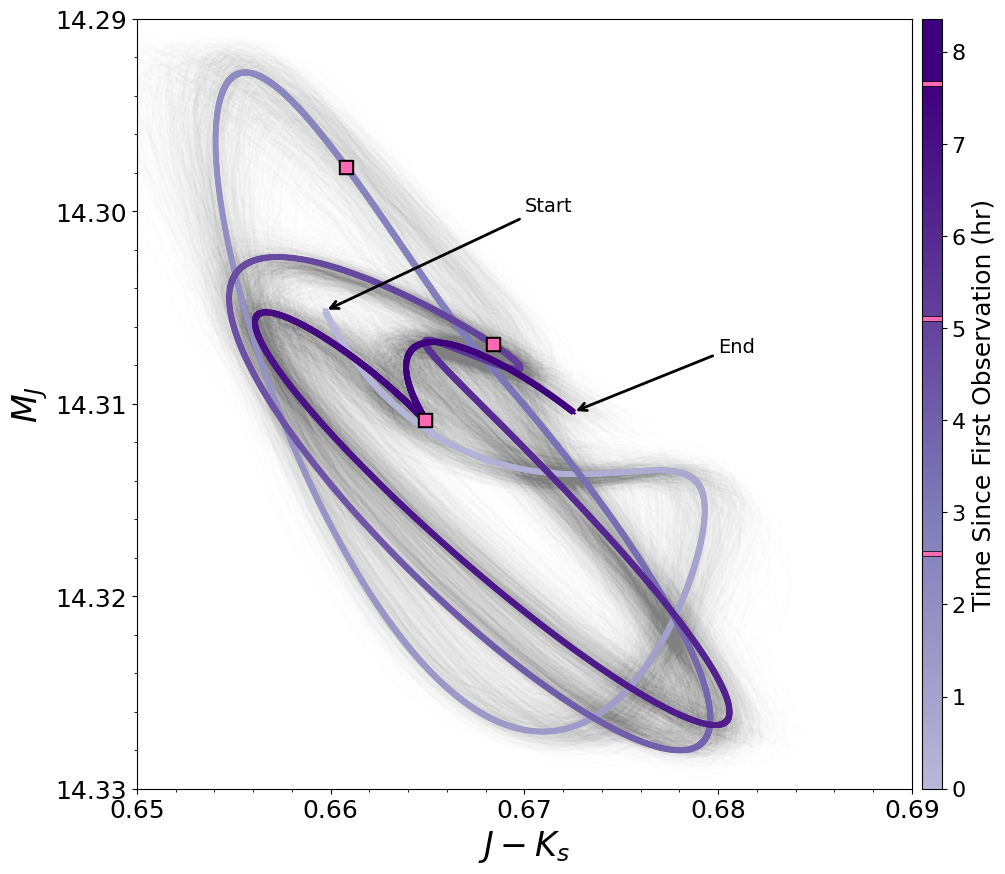}
    \caption{The \textit{J-$K_s$} color difference of the MCMC solutions. The purple line is colored by time since initial observation, where the darkening of the line signals increasing time since initial observation. The gray lines shows the \textit{J-$K_s$} color difference of 1000 random MCMC samples. The pink squares on the color magnitude diagram and pink lines on the color bar show the time of one rotation based on the \textit{J}-band period of 2.55 hr.} 
    \label{fig:ColorChangeSlope}
\end{figure*}

We use the 2MASS transmission curves to determine the cumulative flux in each of the \textit{J-}, \textit{H-} and $K_s$\textit{-}bands, and use the 2MASS zero point flux measurements \citep{2masscal} to convert the measured flux in each band from our three patch atmosphere to a magnitude. We then compared each of the different combined three patch atmosphere colors and magnitudes to our observed data of SIMP0136. In order to convert our SIMP0136 observations to magnitudes, we assume that the normalized flux is equal to the currently accepted \textit{J-}, \textit{H-}, and $K_s$\textit{-}band magnitudes \citep{2021ApJ...911....7Z} which is shown in Figure \ref{fig:JandKLCMags}. The \textit{J-$K_s$} color difference of the MCMC solutions is shown in Figure \ref{fig:ColorChangeSlope}. 

Ideally, simultaneous observations at a wide range of wavelengths would provide better constraints when modeling SIMP 0136’s atmosphere. Previous studies \citep[e.g.][]{2020AJ....159..125L} have found minimal \textit{J-H} color change, $\Delta (J-H) \approx  0$, throughout the rotation of SIMP0136. The lack of color change could be consistent with the conclusions drawn from our contribution plot (Figure \ref{fig:BigPlot}), and the results of \cite{Vos2022}, which place a forsterite cloud between 1.3 and 1.7 bar, above the regions that both \textit{J-} and \textit{H-}band probe, but iron clouds within the region that both \textit{J-} and \textit{H-}band probe. Therefore, any patchiness in either cloud layer would affect both \textit{J-} and \textit{H-}band in the same capacity. Since we do not have \textit{H-}band measurements that are simultaneous or near-simultaneous with our \textit{J-} and $K_s$\textit{-}band measurements, we use the results of \cite{2020AJ....159..125L} and inspect both the \textit{J-$K_s$} and \textit{J-H} color evolution to determine a model's accuracy. To do this, we first match our observed \textit{J-$K_s$} color evolution to a successful model and record the patch ratios that align with the observed data. We do this by taking 30-second steps along the \textit{J-}band and \textit{$J-K_s$} evolution, and determine if there is a point in our model within 0.001 magnitudes from both the \textit{J-}band and \textit{$J-K_s$} magnitudes. From there, any patch coverage combination that was successful along the \textit{J} versus \textit{J-K} evolution is transferred to the \textit{J-H} color space. We are then able to observe the predicted \textit{J-H} behavior. An example is shown in Figure \ref{fig:modelResults}, where the purple line in panel (a) shows the observed data, and the orange stars in panel (a) and panel (b) show the points of successful patch fraction.

\subsection{A Cloudy Patch, Iron-Only Patch, and Clear Patch at 1100K}\label{1100k}

The final scenario discussed in Section \ref{sec:Patchy Cloud Layers}, in which there are three viewing possibilities \textemdash observing a forsterite cloud, observing through a patch in the forsterite cloud layer down to an iron cloud, or observing a patch in both the forsterite and iron cloud layers (clear atmosphere)\textemdash is described in this section. Therefore, we created atmospheres with three patches, all at effective temperatures of 1100K. In each atmosphere, the first patch used one $f_{sed}$ value from the Sonora Diamondback (cloudy) model which includes both forsterite and iron cloud layers. The second patch used one $f_{sed}$ value from the bespoke Sonora Diamondback (iron-only) model, and the third patch used the Sonora Bobcat (clear) atmosphere.

\begin{figure*}
  \centering
  
  \begin{subfigure}{.49\textwidth}
    \centering
    \includegraphics[width=\linewidth]{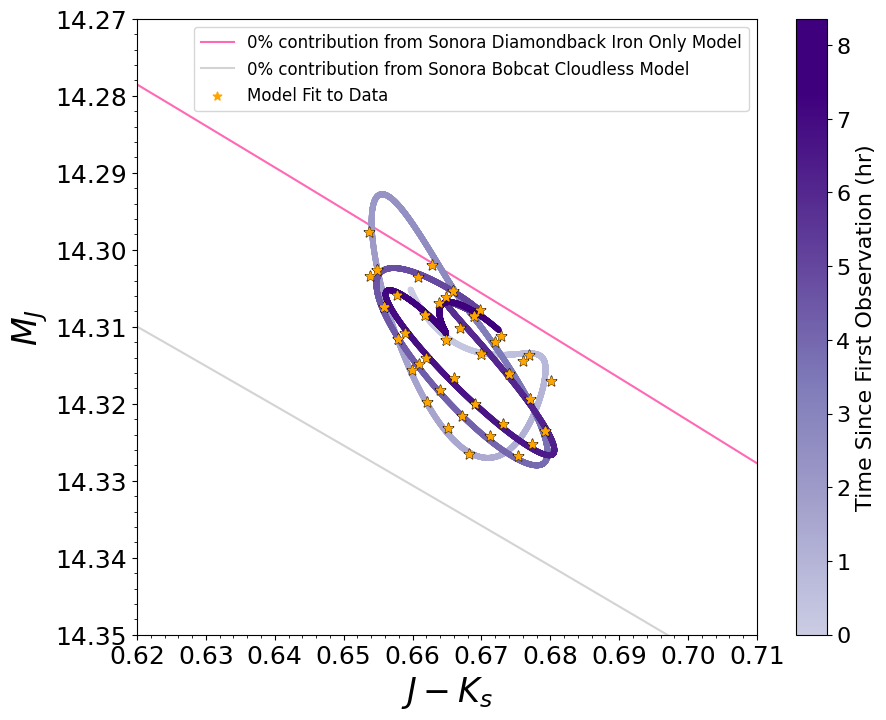} 
    \caption{}
  \end{subfigure}
  \hfill
  \begin{subfigure}{0.49\textwidth}
    \centering
    \includegraphics[width=\linewidth]{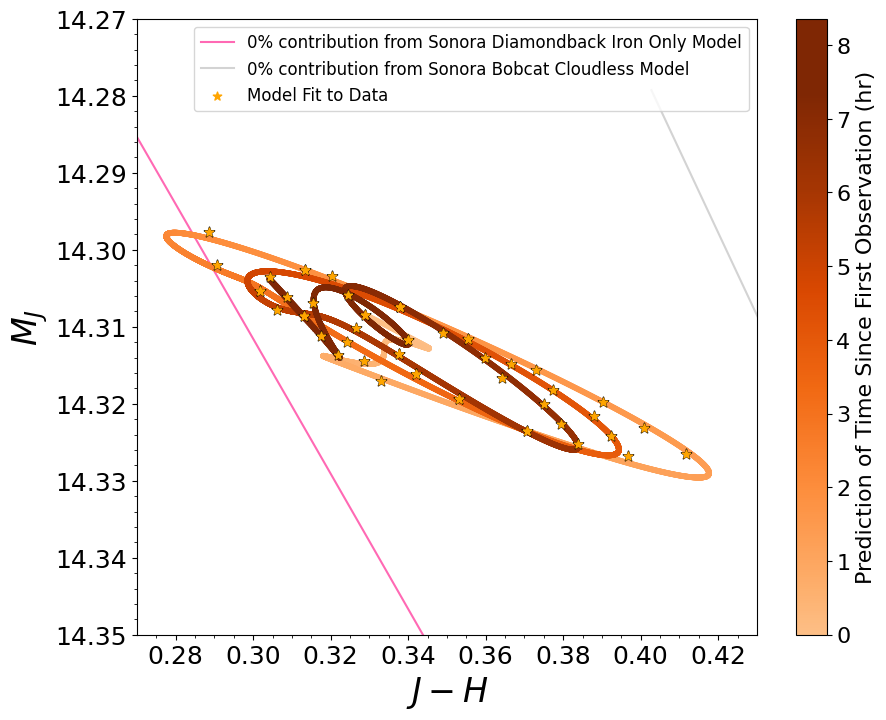}
    \caption{}
  \end{subfigure}
  
  \begin{subfigure}{\textwidth}
    \centering
    \includegraphics[width=\linewidth]{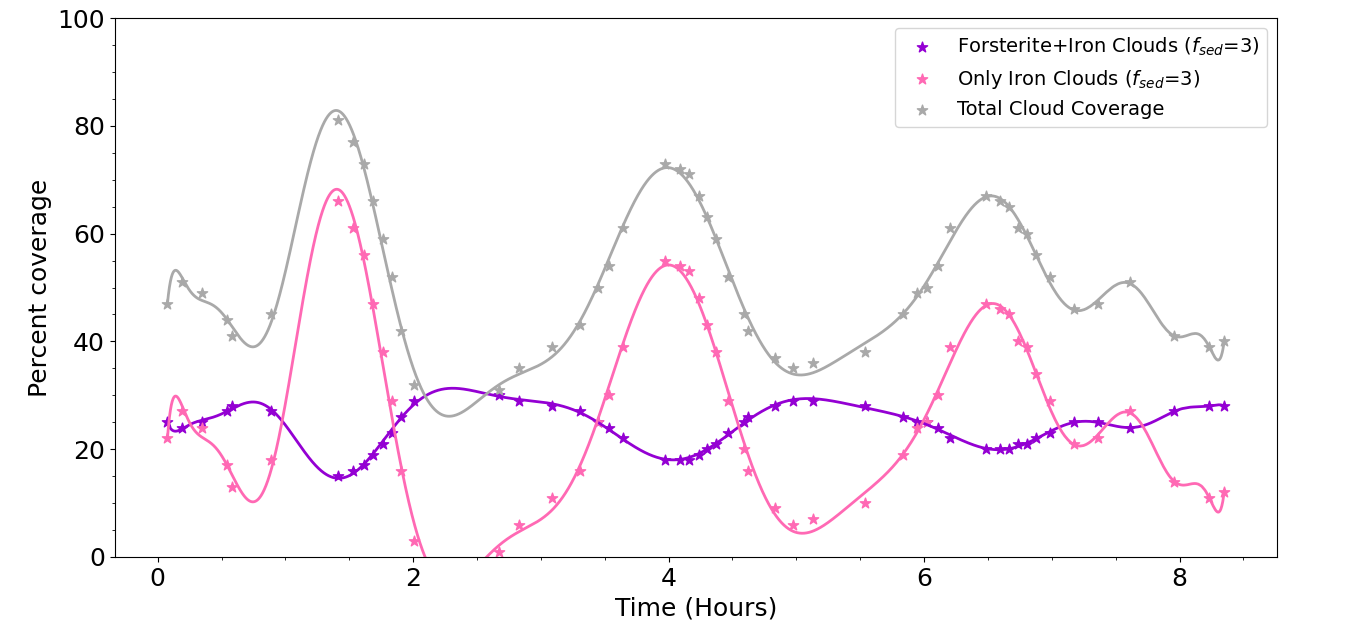} 
    \caption{}
  \end{subfigure}
  
  \caption{A three-patch model where patch 1 uses the Sonora Diamondback models including a layer of forsterite clouds, patch 2 uses the bespoke Sonora Diamondback iron-only models, and patch 3 uses the Sonora Bobcat clear atmospheric models. All three patches have a temperature of 1100K. In this model, both cloudy atmospheres had an $f_{sed}=3$. Panel (a) shows $M_J$ and \textit{$J-K_s$} values from the highest posterior results of the MCMC of our data (Figure \ref{fig:JandKindlc}) in purple. The orange stars represent the points in our model that most closely match our observed data. The fractional coverage ratios that the orange stars represent are then translated into the \textit{J-H} color space, shown in panel (b). In panel (b) the orange line traces the path that the three patch model predicts. In both panel (a) and panel (b),the gray line represents the boundary at which there is 100\% cloud cover (any combination of forsterite and iron), and 0\% contribution by the Sonora Bobcat clear model. The pink line represents the boundary where there is 0\% contribution by the iron only model, so the entirety of the flux is created by the Sonora Bobcat clear model and Sonora Diamondback model which includes forsterite. The corresponding line for when there is 0\% contribution from the Sonora Diamondback which contains forsterite is outside the plotted color-magnitude space, and is thus not included. Panel (c) shows the variation of each patch coverage over time.  The purple stars represent patch 1, the pink stars represent patch 2, and the gray stars show (patch 1 + patch 2) or (1 - patch 3). Panel (c) uses a polynomial fit to trace the change in patch coverage over time. The line is only there as a data visualization tool. From 2.0-2.6 hours, the model cannot reproduce the measured magnitudes, and in panel (c) this is highlighted by the polynomial fit. There is no physical meaning to the fit within this time range.}
  \label{fig:modelResults}
\end{figure*}

In this scenario, multiple model combinations, meaning different combinations of $f_{sed}$ values for patch 1 and patch 2, were able to explain our observed \textit{J-$K_s$} color change. However, when these same models were translated into \textit{J-H} color space, all of the models predicted a significant \textit{J-H} color change throughout the rotation period, conflicting with previous observations. In one such model, shown in Figure \ref{fig:modelResults}, the $\Delta (J-H)=0.12$.

The model shown in Figure \ref{fig:modelResults} uses an $f_{sed}=3$ for both the Sonora Diamondback cloudy atmosphere models, and the bespoke Sonora Diamondback iron-only models. Combining our observations of SIMP0136 with these models, shows that the total fractional cloud coverage would need to vary between 31-81\%, with the forsterite cloud patch varying between 15-30\%, and the iron cloud patch varying between 1-66\%. The forsterite and iron cloud patches are out of phase. This is expected since a patch without forsterite clouds would allow for more iron clouds to be visible, and contribute more to the overall flux.

While this model can explain 93\% of the observed data, the model cannot explain the entirety of the observation (namely the section from 2.2-2.6 hr), and we attribute this to a combination of several factors. First, there may be inaccuracies in the models. Second, an error in the radius measurements of SIMP0136, which are needed to convert the model flux values to magnitudes, and third, an error in the magnitudes arising from using the measured 2Mass values as a reference point. At the peak variability of the \textit{J-}band light curve, from the trough at $\sim$ 1.47 hr to the peak at $\sim$ 2.37 hr, the amplitude is $3.13 \pm 0.16$\%. The cloud coverage at the trough is $\sim$ 15\% forsterite+iron and $\sim$ 68\% iron only, corresponding to $\sim$83\% complete cloud cover. At the time of the peak, our data is outside of the model. Our best approximation would be $\sim$ 31\% forsterite+iron clouds, and $\sim$ 0\% iron only cloud coverage. This corresponds to an increase in the forsterite+iron cloud coverage of $\sim$ 16\% decrease in iron cloud coverage of $\sim$ 68\%, and a decrease in the total cloud coverage of $\sim$ 52\%.

As mentioned, other model combinations were successful at describing our observed \textit{J-$K_s$} color change. Interestingly, there is little to no change when changing either patch 1 or patch 2's $f_{sed}$ from $f_{sed}=3$ to $f_{sed}=4$. However, when changing either patch to $f_{sed}=1$ or $f_{sed}=2$, while the other patch remains a higher $f_{sed}$, the models can no longer match the data. Once both patches have an $f_{sed}=1$ or $f_{sed}=2$, the model can match the data again. When the patches have a lower $f_{sed}$, the overall cloud coverage drops to a maximum of 35\%. The forsterite and iron clouds are still out of phase, and both species coverage fraction fluctuates $\sim$30\%.

\subsection{Inhomogenous Surface Temperature Patches for Both Entirely Clear and Entirely Cloudy Atmospheres}

For the sake of completeness, in this section we consider whether a three-temperature model can also explain the observed phase shift in \textit{J-} and \textit{$K_s$}-band, and lack of phase shift between \textit{J}- and \textit{H}-band.

\subsubsection{Clear Atmosphere}\label{521}

\cite{2014ApJ...785..158R} found that atmospheric temperature fluctuations could be responsible for observed photometric variability in brown dwarfs.  

Using the Sonora Bobcat (clear) atmospheric models, we found that in order to create a red enough atmosphere to match our observed range in \textit{J-$K_s$}, we required three patches at effective temperatures of 1500K, 900K, and 800K. In this scenario, our $\Delta (J-H)=0.03$, which is still different than the previous \textit{J-H} observations of SIMP0136.

\subsubsection{Cloudy Atmosphere}\label{522}

Using different effective temperature patches of the standard Sonora Diamondback (cloudy) models, we are unable to replicate the atmosphere of SIMP0136. The spectra are too red to match the observed range in \textit{J-$K_s$}. It should be noted that as described in Section \ref{sec:modelingatmosphere}, our \textit{J-} and $K_s$\textit{-}band magnitudes were based on the currently accepted values for SIMP0136. It is possible that at the time our observations were completed, the \textit{J-} and $K_s$\textit{-}band magnitudes were different than at the time the currently accepted values were measured.

\subsection{A Combination of Clear and Cloudy Patches at Different Temperatures}

Finally, we used patches of different species of clouds, at different $f_{sed}$ values, at different effective temperatures, to attempt to describe the observed \textit{J-$K_s$} color change and lack of \textit{J-H} color change. 

\subsubsection{Two Clear Patches at Different Temperatures, with a Third Patch which is Cloudy}\label{531}

Multiple models of combining two clear patches at different effective temperatures with a third cloudy patch were able to explain the observed \textit{J-$K_s$} color change. However, none of them are able to explain the previously observed \textit{J-H} color change. For one such case, where patch 1 was a clear patch at $T_{eff}=$1300K, patch 2 was a clear patch at $T_{eff}=$1200K, and patch 3 was a cloudy patch at $T_{eff}=$900K and $f_{sed}=4$, the $\Delta (J-H)=0.02$.

\subsubsection{Two Cloudy Patches at Different Temperatures, with a Third Patch which is Clear}\label{532}

Many different combinations of two cloudy patches with a third clear patch at different effective temperatures were able to explain the observed \textit{J-$K_s$} color change but not the previously observed \textit{J-H} color change. For a clear patch at $T_{eff}=$1200K, and cloudy patches at $T_{eff}=$900K and $T_{eff}=$1100K, both with $f_{sed}=3$, the resulting $\Delta (J-H)=0.02$. 

\subsubsection{A Cloudy Patch, Iron-Only Patch, and Clear Patch, all at Different Temperatures}\label{533}

Several different combinations of effective temperatures and $f_{sed}$ values were able to explain the observed \textit{J-$K_s$} color change, but unable to explain the lack of \textit{J-H} color change. For a model atmosphere with a clear patch at $T_{eff}=$1000K, an iron-only patch with an effective temperature of $T_{eff}=$1100K and $f_{sed}=3$, and a cloudy patch with an effective temperature of $T_{eff}=$1200K and $f_{sed}=4$, the $\Delta (J-H)=0.01$. This scenario represents our most successful attempt at matching both the observed \textit{J-$K_s$} color change, as well as the previously observed lack of \textit{J-H} color change. As mentioned in the introduction, \cite{2020AJ....159..125L} does not mention specific values for the $\Delta (J-H)$ color change, but based on the appearance of the linear fit in Figure 5, we can only approximate it to $\Delta (J-H)=0$ with zero uncertainty. Without having simultaneous \textit{J-H} measurements of our own, we are unable to say with certainty that this model accurately represents the atmosphere of SIMP0136.

\section{Conclusion}

In this work, we present the first detected \textit{J-} and $K_s$\textit{-}band phase shift measurement for SIMP0136, a planetary-mass T2.5 dwarf. 
We determine the phase shift to be $39.9^{\circ +3.6}_{ -1.1}$. The period of the light curve is found to be $2.55^{+0.10}_{-0.08}$ and $2.68 \pm 0.28$ hours, for \textit{J-} and $K_s$\textit{-}band respectively. The \textit{J-} and $K_s$\textit{-}band periods are consistent to within 1$\sigma$. Both the \textit{J-} and \textit{$K_s$-}band light curves have amplitudes which decrease over time. The \textit{J-}band amplitude decreases from  $3.13 \pm 0.16\%$ to $2.27 \pm 0.10\%$ to $1.93 \pm 0.12\%$. For $K_s$\textit{-}band, the amplitude decreases from $1.94 \pm 0.27\%$ to $1.57 \pm 0.19\%$ to $1.17 \pm 0.16\%$.

Additionally, using the Sonora Bobcat models \citep{Sonora}, we calculated the flux distribution of a cloudless model atmosphere for SIMP0136 as a function of pressure and wavelength (Figures \ref{fig:BigPlot}, \ref{fig:F_P}), and find that {the \textit{J-}, \textit{H-}, and $K_s$\textit{-}band primarily probe the pressures 12.6\textendash 26.2, 3.6\textendash 20.0, and 1.9\textendash 14.1 bar, respectively.} 

We find that among the explored various non-homogeneous temperature and cloudy scenarios tested in this paper multiple patchy cloud layers is the preferred model in order to describe the phase shift detected between the \textit{J-} and $K_s$\textit{-}band light curves.
The two patchy layers include a patchy upper layer, which sits within or above the region of the atmosphere that the $K_s$\textit{-}band probes, and a patchy lower cloud layer that sits either between the regions of the atmosphere which \textit{J-} and $K_s$\textit{-}band probe or within the region of the atmosphere that the \textit{J-}band probes (final scenario in Section \ref{sec:Patchy Cloud Layers}). This expands on the results of \cite{Vos2022}, who find that the spectrum of SIMP0136 is best described by a model containing patchy forsterite clouds at a pressure range of 1.3\textendash 1.7 bar above an iron cloud deck sitting at and below $\sim$7 bar. 

We attempted to model the atmosphere of SIMP0136 by combining multiple spectra as different patches representing fractional coverage in the atmosphere. We used the Sonora Bobcat (clear) atmospheric models, the Sonora Diamondback (cloudy) atmospheric models containing forsterite and iron clouds, and bespoke Sonora Diamondback models which contain only iron clouds. In every combination of model we could create, we were unable to match both the observed \textit{J-$K_s$}=0.03 color change and previously measured lack of \textit{J-H} ($\Delta J-H \approx 0$) color change from \cite{2020AJ....159..125L}. 

The combination of different spectra is an approximation of the potential atmospheric structure of SIMP0136. In every combination, the models are not able to replicate the previously observed lack of \textit{J-H} color change from \cite{2020AJ....159..125L}. However, these observations are not simultaneous or near-simultaneous with the \textit{J-} and $K_s$\textit{-}band observations presented in this work, and it is unclear how much of a problem this poses. If the minimal \textit{J-H} color change is a constant feature, this suggests that the atmosphere of these objects is much more complex, and our understanding of the cloud layers between 1 and 20 bar is incomplete. Moving forward, simultaneous or near-simultaneous observations of the \textit{J-}, \textit{H-}, and $K_s$\textit{-}band will be necessary to gather information about the atmosphere of these objects, and the atmospheric layers that the near infrared bands probe.

Forthcoming JWST observations will reveal far more about the atmospheres of substellar objects given that multiple programs will be monitoring objects over extended wavelength coverage.Long term monitoring is necessary to determine if phase shifts, amplitudes, and color-changes are constant overtime. Simultaneous multiwavelength photometry combined with the power of atmospheric models, such as the Sonora Bobcat and Diamondback models, but also 3D atmospheric modeling, provide an opportunity to create a more complete map of the atmospheric structure of exoplanet analogs, such as SIMP0136.

\begin{acknowledgments}
We thank the reviewer for their helpful comments, which improved the quality of this publication. 
We also thank Michael Line for his insight into the use of atmospheric models, such as Sonora Bobcat. 

This material is based upon work supported by the National Aeronautics and Space Administration under grant No. 80NSSC20K0256 issued through the Science Mission Directorate. 

This material is based upon work supported by the National Science Foundation Graduate Research Fellowship under Grant No. DGE-1840990.

J. M. V. acknowledges support from a Royal Society - Science Foundation Ireland University Research Fellowship (URF$\backslash$1$\backslash$221932).
\end{acknowledgments}

\vspace{5mm}

\software{astropy \citep{astropy:2022},  
          SciPy \citep{SciPy-NMeth2020}, 
          Celerite2 \citep{celerite2},
          emcee \citep{2013PASP..125..306F}
}

\bibliography{bibliography}{}
\bibliographystyle{aasjournal}

\end{document}